**TABLE 1**
Limiting Magnitudes of GWS and MDS Fields

| Survey | $l$ [deg] | $b$ [deg] | $N_i$ | $N_{field}$ | $t_i$ [s] | $sky_I$ [hr$^{-1}$] | $I_{lim}$ [mag] |
|--------|-----------|-----------|-------|-------------|-----------|---------------------|------------------|
| GWS | 96.3 | 60.2 | 12 | 1 | 25200 | 6.0 | 25.3 |
| GWS | 96.3 | 60.2 | 4 | 27 | 4400 | 6.0 | 24.3 |
| MDS | 215.2 | −87.6 | 2 | 1 | 4200 | 14.6 | 24.4 |
| MDS | 83.9 | −76.4 | 2 | 1 | 2000 | 30.2 | 23.6 |
| MDS | 133.9 | −64.9 | 3 | 1 | 6300 | 40.3 | 24.0 |
| MDS | 303.3 | −65.0 | 2 | 1 | 2000 | 11.2 | 24.0 |
| MDS | 233.7 | −62.4 | 2 | 1 | 4200 | 14.1 | 24.3 |
| MDS | 123.7 | −50.3 | 3 | 1 | 6300 | 44.3 | 23.9 |
| MDS | 123.8 | −50.0 | 2 | 1 | 4200 | 16.7 | 24.4 |
| MDS | 178.5 | −48.1 | 5 | 1 | 6600 | 7.7 | 24.4 |
| MDS | 179.8 | −32.2 | 7 | 1 | 5900 | 13.9 | 23.9 |
| MDS | 326.4 | −29.6 | 3 | 1 | 6300 | 8.8 | 24.6 |
| MDS | 82.0 | −19.2 | 3 | 1 | 4800 | 7.5 | 24.6 |
| MDS | 81.8 | −19.2 | 2 | 1 | 4200 | 11.5 | 24.6 |
| MDS | 43.7 | 20.3 | 3 | 1 | 6300 | 7.0 | 24.7 |
| MDS | 206.1 | 19.6 | 2 | 1 | 4200 | 55.7 | 23.8 |
| MDS | 52.1 | 27.8 | 4 | 1 | 2280 | 7.4 | 24.0 |
| MDS | 52.0 | 27.8 | 2 | 1 | 2000 | 14.1 | 24.0 |
| MDS | 303.4 | 33.6 | 4 | 1 | 8400 | 6.8 | 24.8 |
| MDS | 56.7 | 34.2 | 3 | 1 | 6300 | 6.1 | 24.8 |
| MDS | 206.8 | 35.7 | 2 | 1 | 6900 | 19.7 | 24.3 |
| MDS | 16.0 | 39.9 | 6 | 1 | 6000 | 5.5 | 23.7 |
| MDS | 16.2 | 40.0 | 12 | 1 | 12000 | 30.1 | 24.6 |
| MDS | 35.8 | 56.5 | 4 | 1 | 7500 | 7.8 | 24.6 |
| MDS | 35.6 | 56.4 | 3 | 1 | 6000 | 16.5 | 24.4 |
| MDS | 359.0 | 64.7 | 4 | 1 | 6900 | 13.9 | 24.4 |
| MDS | 34.0 | 66.7 | 6 | 1 | 6000 | 4.4 | 24.4 |
| MDS | 34.4 | 66.6 | 8 | 1 | 8000 | 3.1 | 24.7 |
| MDS | 202.3 | 76.4 | 3 | 1 | 6300 | 10.8 | 24.6 |
| MDS | 290.9 | 74.4 | 2 | 1 | 2700 | 28.1 | 23.9 |

# EVIDENCE FOR GALAXY INTERACTIONS/MERGERS
# FROM MEDIUM-DEEP SURVEY WFPC2 DATA[1]

L. W. Neuschaefer[2], M. Im, K. U. Ratnatunga, R. E. Griffiths, Casertano, S.

Bloomberg Center for Physics and Astronomy, Johns Hopkins University, Baltimore, MD 21218;

lwn, myung, kavan, griffith@mds.pha.jhu.edu, stefano@stsci.edu

## ABSTRACT

We examine the morphological and statistical properties of close galaxy pairs from two sets of 28 WFPC2 fields, acquired for the Medium Deep Survey (MDS) and the Groth/Westphal Survey (GWS) in F606W ($V$) and F814W ($I$) passbands. In the GWS sample all fields have uniform 95% completeness down to $I \lesssim 24.3$ mag, whereas in the MDS sample, fields have varying 95% completeness limits in the range $I \lesssim 23.6$–24.8 mag. In each field $\sim 400$ galaxies per $\simeq 5$ square arcminute field are detected. We exploit high resolution WFPC2 images to systematically determine morphological classifications of galaxies as disk- or bulge-dominated down to $I \leq 23$ mag ($V \lesssim 24$ mag), and to differentiate galaxies from stars one magnitude fainter. Down to $I \leq 25$ mag the number of galaxy pairs with separations $\theta \leq 3''.0$ is consistent with a shortward extrapolation of the angular two-point correlation function $\omega(\theta) \propto \theta^{-0.8}$ observed from the same data; the fraction of such pairs showing morphological evidence for physical association accounts for a third of the total numbers suggested by a shortward extrapolation of $\omega(\theta)$. The latter result may not be too surprising given the low surface brightness of the tidal tails resulting from galaxy interactions, i.e. much of the evidence for interactions may fall below our detection limit. Moreover, we find no trend between apparent physical association (on the basis of morphology) and ($V-I$) color or $I$-magnitude difference between pair members of the $\theta \leq 3''.0$-pair sample.

We use recent galaxy redshift surveys to estimate the rate of galaxy merging occurring in the MDS and GWS galaxy pair samples. From this work we find that merging has a moderate dependence on redshift: we derive an estimate for the galaxy pair fraction $P_{\rm f} \propto (1+z)^m$, with $m = 1.2 \pm 0.4$ for galaxies with $I \leq 25$ mag ($z_{\rm med} \lesssim 1$–2). Two scenarios are consistent with both this low value of $m$ and for the low correlation amplitude: (a) a low density universe with strong clustering evolution parameterized by a clustering exponent $\epsilon \simeq 1.0$ such that galaxy cluster-scale structures shrink relative to the proper coordinate frame; and/or (b) a weakly clustered galaxy population, the majority of which fade or dissipate below $z_{\rm med} \lesssim 0.5$ ($I \lesssim 20$ mag), thus mimicking the apparently strong evolution in the correlation amplitude, $A_{\omega}$. Although not directly observed using our data, the possible flattening of the slope of $\omega(\theta)$ with increasing survey depth can explain the strong decline in $A_{\omega}$ and allow for greater pair fraction evolution limited to $m \leq 1.6$.

*Subject headings:* galaxies: evolution – galaxies: clustering – galaxies: structure – cosmology: observations – large-scale structure of the universe

---

[2] Current address: Bloomberg Center for Physics and Astronomy, The Johns Hopkins University, San Martin Dr., Baltimore, MD 21218-2695; email lwn@mds.pha.jhu.edu.

# 1. INTRODUCTION

There have been numerous attempts to reconcile the apparent evolutionary effects suggested by the excess galaxy number counts for $B_J \gtrsim 22.5$ mag (Tyson 1988; Maddox et al. 1990a; Lilly, Cowie & Gardner 1991; Metcalfe et al. 1993) with the apparent lack of evolutionary effects found by interpretation of the $K$-band galaxy number counts (Gardner, Cowie & Wainscoat 1993) and the galaxy redshift distribution, $N(z)$ for $B_J \lesssim 22.5$ mag (Broadhurst, Ellis & Shanks 1988; Colless et al. 1990, 1993; Lilly et al. 1995). The explanation for the fate of the galaxies responsible for the excess blue number counts can possibly be constrained by the degree to which they are observed to be merging with their neighbors. The high resolution WFPC2 images studied in this paper provide direct evidence for the frequency of galaxy merger events down to $I \leq 25$ mag (or median redshifts $z_{\mathrm{med}} \lesssim 1\text{–}2$). With this information, we are able to shed new light on the nature of the faint blue galaxies, whose origin has the following two extreme interpretations: (a) they are members of a low mass population which, by the present epoch, has either faded below present detection limits (Tyson 1988) or dissipated owing to internal supernova driven winds (Babul & Rees 1992); or (b) they have taken part in wholescale merging (Guiderdoni & Rocca-Volmerange 1991; Broadhurst, Ellis & Glazebrook 1992).

The angular two-point correlation function, $\omega(\theta)$, has a bearing on this issue. Over the past several years $\omega(\theta)$ has been measured numerous times within the range $22 \lesssim B_J \lesssim 26$ mag (Efstathiou et al. 1991; Roche et al. 1993; Couch, Jurcevic & Boyle, 1993; Infante & Pritchet 1995; Neuschaefer & Windhorst 1995 (NW95); Neuschaefer et al. 1995a; Brainerd, Smail & Mould 1995). By convention a power law has been used to describe $\omega(\theta)$:

$$\omega(\theta) = A_\omega \theta^\delta \tag{1}$$

thereby defining the correlation amplitude, $A_\omega$, and slope, $\delta$. All the above authors find a strong decline in $A_\omega$ with increasing survey depth, together with moderate (NW95) to non-existent flattening (Couch, Jurcevic & Boyle 1993) in the correlation slope when compared with the locally determined value of $\delta \simeq -0.8$ (Groth & Peebles 1977). Clustering evolution in which $A_\omega$ is parameterized as:

$$A_\omega \propto (1+z)^{\gamma - 3 - \epsilon} \tag{2}$$

can match the observations using $\epsilon \sim 1$ in a flat $q_0 = 0.5$ universe without biased galaxy formation. This model is consistent with the linear growth of density perturbations in the Cold Dark Matter scenarios of Yoshii, Peterson & Takahara (1993), in which galaxy clusters shrink in scale relative to the *proper* coordinate frame. The CDM model simulated by Yoshii et al. (1993) also shows that $\omega(\theta)$ might tend to flatten toward smaller angular scales.

Interpretation (a) has been argued by Efstathiou et al. (1991), who concluded in their study of $\omega(\theta)$ that faint blue galaxies are weakly clustered. They noted that, by virtue of their color, such blue galaxies were likely to be of relatively late morphological type, and these were known to be more weakly clustered than E/S0 galaxies, as observed in local surveys (Giovanelli, Haynes & Chincarini 1986). This conclusion was bolstered by observations of $A_\omega$ versus broad-band colors: $A_\omega$ is factors of several larger for red-selected as opposed to blue-selected samples (Bernstein et al. 1994; Landy, Szalay & Koo 1996). Moreover, down to $B_J \leq 22.5$ mag, Bernstein et al. (1994) affixed a physical scale to their photographic plate data by replicating the observing conditions of the redshift survey of Colless et al. (1993).

From this they inferred that the bluest two-thirds of their sample have clustering properties similar to IRAS galaxies observed locally (Saunders et al. 1992), i.e. they are $\sim$2–3 times less clustered than near-$L^*$ galaxies comprising the APM catalog (Maddox et al. 1990b).

Interpretation (b) has been argued by Broadhurst, Ellis & Glazebrook (1992), who constructed a merger model consistent with the deepest available $B_J$- and $K$-band number counts and a $B_J$-selected redshift distribution. In their model, the decline in $A_\omega$ with increasing sample depth is related to the merger process, although it is not clear what effect this has on the shape of $\omega(\theta)$ on small scales. Depending on the merger timescale, $\tau_{\mathrm{mg}}$, the slope of $\omega(\theta)$ may steepen, reflecting the merging pairs in excess of the underlying pair distribution.

The morphologies and colors of close galaxy pairs may offer an indicator as to whether or not the pair members are physically associated. Numerical simulations of galaxy merging at intermediate redshift indicate that it may be very rare to observe merging in process; after a few tenths of a Gyr only very low surface brightness tidal debris may be left to indicate the occurrence of a merger event (Mihos 1995). Using such morphological evidence, as well as broad-band colors and statistics of the fraction of close galaxy pairs, $P_{\mathrm{f}}$, the evolutionary index for the redshift dependence of $P_{\mathrm{f}}$ has been estimated to fall in the range $P_{\mathrm{f}} \propto (1+z)^m$, $2.4 \lesssim m \lesssim 4$ (Zepf & Koo 1989; Burkey et al. 1994; Carlberg, Pritchet & Infante 1994; Yee & Ellingson 1995). These results are supported by a study using relative line-of-sight pairwise velocities of 14 galaxies in $\theta < 6''$ pairs (Carlberg et al. 1994). However, not all studies find significant evidence for such a high evolutionary index based on statistics of close galaxy pairs. On the contrary, Woods, Fahlman & Richer (1995) find evidence for only a small evolution in the merger rate. Such discrepancies may be a reflection of the sparse statistics available for some of these studies, especially the spectroscopic ones, and possibly real field-to-field differences in small scale clustering phenomena.

To help resolve the origin of the faint blue galaxies (see also Griffiths et al. 1994; Glazebrook et al. 1995; Driver, Windhorst & Griffiths 1995) and to assess the importance of galaxy merging in their interpretation, we examine the morphology, colors and statistics of close galaxy pairs from two deep, complementary field galaxy samples using WFPC2. Using a very large sample of galaxies classified completely down to $I \lesssim 23.5$ mag we have the capacity to test for field-to-field fluctuations in galaxy pairs and to better constrain the rate of galaxy mergers. In §2 we describe the images, the sky coverage and the derived catalogs. In §3 we place constraints on the rate of galaxy mergers versus redshift from nearest neighbor statistics. In §4 we examine the morphological properties of close galaxy pairs in greater detail. In §5 we discuss our results. Finally in §6 we present our conclusions. Unless specifically stated otherwise, in this paper we assume $H_0$= 100 km/s/Mpc, $q_0$=0.5.

## 2. CATALOGS

The catalogs used for the present study are derived from WFPC2 direct images obtained for the Medium Deep Survey (MDS) and the Groth/Westphal Survey (GWS) in $F606W$ ($V$) and $F814W$ ($I$) filters. Both samples are comprised of 28 fields, with the MDS sample uniformly distributed over the sky with $|b| \geq 20°$, and the GWS sample arranged in a contiguous linear strip at $l$=96°.3, $b$ = +60°.2. Table 1 provides information on the location and depth of each field in the $I$-band. With the exception of one ultra-deep field of 25,200s, each of the GWS fields had total integration times of 4400s, whereas MDS fields had integration times in the range 2000–12,000s. The completeness of each field depends on contributions from the sky background and read-noise, as well as on galaxy magnitude and surface-brightness. Column eight of Table 1 gives the nominal 95% completeness limit for galaxies with half light radii $r_{hl} \leq 0''.8$. The zeropoint for these completeness estimates are determined via Monte Carlo modeling of the detection process using simulated galaxy images incorporating the optical and noise properties of WFPC2 and the HST optical telescope assembly; for further details see Neuschaefer et al. 1995b. The two samples are complementary since the MDS sample includes diverse lines of sight over which to obtain a relatively unbiased measure of small-scale clustering, whereas the GWS sample spans both small and intermediate angular scales, and so provides a measure of small- to intermediate-scale clustering with relatively lower internal dispersion. Each field subtends a solid angle of 5.1 arcmin$^2$ per pointing and typically contains ∼400 objects per field.

Our detection software is optimized to resolve object pairs with separations $\theta \gtrsim 0''.5$. Detected objects are classified on the basis of morphology after the removal of systematics. Parameter estimates for (x,y) position, flux, image profile (e.g., point source, disk-like or bulge-like), half-light radius, position angle and axial ratio are obtained using the following modelling method. From a trial galaxy model an image is constructed, convolved by a model WFPC2 point spread function (using the software package tinytim v.4b; see Krist 1995), and spatially integrated to the WFPC2 pixel scale of $0''.1$. The model image is then compared with the observed galaxy image. Assuming each pixel in the observed image has a Gaussian error distribution with respect to the model, we derive the probability for obtaining the observed pixel value. We define the likelihood function as the sum of the logarithms of such probabilities, which is similar to a weighted $\chi^2$. This likelihood function is then minimized using a quasi-newtonian method (Ratnatunga, Griffiths & Casertano 1994).

Over 70% of all detected objects are classified as galaxies. Figure 1 shows the differential number counts for objects classified as predominantly disk-like ('disks'), predominantly bulge-like ('bulges') and generic galaxies in the GWS Survey. The generic galaxy classification applies to objects with extended profiles for which our software does not find a significantly superior fit among the disk and bulge models. The limiting magnitude for complete classifications corresponds to the plateaus in the disk and bulge counts around $I \simeq$ 23.5 mag and $I \simeq 23$ mag, respectively.

At the time of this analysis we had access to a preliminary version of the automated catalogs for which ∼10% of the objects have suspicious fits, as measured by substantial discrepancy between the total flux estimated from the best fitting model, $I_{mod}$, and that within the one sky-sigma isophote, $I_{iso}$. We visually inspected objects with $I_{iso} - I_{mod} > 1$ mag; as a result of this effort $\simeq 5\%$ of the catalog objects were reclassified and $\simeq 4\%$ were discarded as spurious. For each reclassified object the errant model magnitude was replaced by the corresponding isophotal magnitude, removing a small zeropoint difference of −0.2 mag between the mean model and mean isophotal magnitudes.

## 3. OBSERVATIONS

### 3.1 Angular Correlation Functions $\omega(\theta)$

The two-point correlation function $\omega(\theta)$ is implicitly defined via the probability, $dP$, of simultaneously finding galaxies within the solid angles $d\Omega_1$ and $d\Omega_2$, separated by an angle $\theta$, in a survey region with galaxy surface density $n$:

$$dP(\theta) = n^2(1 + \omega(\theta))d\Omega_1 d\Omega_2 \tag{3}$$

We estimate the angular two-point correlation function, $\omega(\theta)$, using the estimator $\hat{\omega}(\theta)$ discussed by Landy & Szalay (1993):

$$\hat{\omega}_f(\theta) = \frac{n_{dd}(\theta) - 2n_{dr}(\theta) + n_{rr}(\theta)}{n_{rr}(\theta)} \tag{4}$$

where $n_{dd}$, $n_{dr}$ and $n_{rr}$ are the data-data, data-random and random-random pairs counted at separation $\theta \to \theta + d\theta$. Eq.(3) suffices to estimate $\omega(\theta)$ for galaxy samples extracted from the single GWS field, whereas for the 28 separated MDS fields we compute the average $\hat{\omega}(\theta)$ as a sum of the individual field $\hat{\omega}(\theta)$'s, weighting by the number of galaxies $N_i$ in the $i$th field:

$$\hat{w}(\theta) = \frac{\Sigma N_i \hat{\omega}_i}{\Sigma N_i}. \tag{5}$$

As an alternative, $n_{dd}$, $n_{dr}$ and $n_{rr}$ can be summed over the 28 MDS fields, from which an estimator for $\omega(\theta)$ can be obtained via eq.(4). Relative to the estimator of eq.(5), this alternative estimator of $\omega(\theta)$ has an rms difference of $\lesssim 8\%$, with neglible mean difference. The random errors in the estimator $\hat{\omega}(\theta)$ are dominated by $n_{dd}$ since the random fields that we use to normalize $\hat{\omega}(\theta)$ are typically factors of 10–20 larger than a given galaxy sample. Thus, $\hat{\omega}(\theta)$ for the single contiguous GWS field has smaller random errors than those for the sum of the 28 separated MDS fields. Relative to the sum of the individual MDS fields, the single GWS field has factors $\simeq 28$ and $\simeq 2.5$ more galaxy *pairs*, out to the full $43'$ extent of the GWS field and within the $\theta \lesssim 3'\!.5$ extent of a single WFPC2 frame, respectively.

The normalization in eq.(3) induces a negative bias in $\hat{\omega}(\theta)$ because of an integral constraint discussed by Peebles (1980). This correction arises through the implicit assumption of eq.(3) that the integral of $\hat{\omega}(\theta)$ over the survey region vanishes. For small survey regions, such as the individual WFPC2 frames of the MDS sample, and to a lesser extent the contiguous GWS sample, $\hat{\omega}(\theta)$ is always positive. For example, at a scale of one arcminute, the correction due to the integral constraint amounts to $\sim 50\%$ and $\sim 15\%$ for individual WFPC2 frames and the contiguous GWS survey region, respectively. We correct for the integral constraint and for higher order correlations using Equation (27) of Bernstein (1994).

Figure 2 shows $\omega(\theta)$ for galaxies in the GWS field. Disk-like, bulge-like and generically classified galaxies are included in the samples. For improved visibility the top and middle correlation functions have been offset by 4 and 2 dex, respectively. All three magnitude intervals show that $\omega(\theta)$ conforms roughly to a power law with slope $\delta \simeq -0.8$ as observed in local surveys. The aforementioned corrections to $\omega(\theta)$ have been applied to the individual data points. Poissonian errors are shown; the lower limit for many of the data points allow for $\omega(\theta) < 0$, as indicated by the downward pointing arrows. Downward pointing arrows without open squares indicate only upper limits for those points with *corrected* values of $\omega(\theta) < 0$.

The correlation amplitudes, $A_\omega$, for disk, bulge and combined galaxy samples are shown in Figure 3 and listed in Table 2 for the GWS and MDS samples in the upper and lower panels, respectively. To determine $A_\omega$ we used angular scales $1'' \leq \theta \leq 1'\!.2$ and $1'' \leq \theta \leq 10'$ for the MDS and GWS samples, respectively. We used non-linear parameter estimation to fit for $A_\omega$, following an approach outlined by Bernstein (1994). Because of the varying levels of completeness of the MDS fields (see Table 1), shallow MDS fields are excluded so as to estimate amplitudes for nominally *complete* galaxy samples down to $I_{med} \leq 24$ mag for *both* the MDS and GWS samples. A caveat applies for amplitudes with $I_{med} \geq 24$ mag in both MDS and GWS samples, however, because of the preferential selection of high-surface brightness galaxies at the faintest magnitudes.

Because of the integral constraint bias and because $A_\omega$ and $\delta$ are zeroth and first order fitting parameters to $log\ \omega(\theta)$ versus $log\ \theta$, respectively, the slope $\delta$ is more sensitive than $A_\omega$ to possible biases in the integral constraint correction. Thus we use the large-scale GWS galaxy sample to minimize this effect to obtain a relatively unbiased measurement of $\delta$. Using the same non-linear fitting method used to estimate $A_\omega$, we find the slopes of the correlation functions in Figure 2 to be $\delta = -0.86 \pm 0.12$, $-0.75 \pm 0.10$ & $-0.8 \pm 0.2$, top to bottom (measured over the range $10'' \leq \theta \leq 5'$). Within measurement errors $\omega(\theta)$ maintains a constant slope, but this does not preclude the possible flattening of $\omega(\theta)$ which has been used as means to explain the strong decline in $A_\omega$ (NW95).

All the models plotted in Figure 3 are normalized with the clustering scale length of $r_0 = 5.5h^{-1}Mpc$ which has been found to be appropriate for local $L^*$ galaxies (Davis& Peebles, 1983), and is in agreement with the $B_J$ correlation amplitudes of Maddox et al. (1990) after transformation to $F814W$. We consider the model first proposed by NW95 for evolution of the slope, parameterizing $\gamma \equiv 1 - \delta = (1 + z)^{-C}$, with $C = 0.2$. The dotted curve lies below the correlation amplitudes for the full galaxy samples in the GWS survey, but is in good agreement with the amplitudes for the MDS full galaxy sample. This model can account for the strong decline in $A_\omega$, but assumes marginal evolution in the slope. The latter is consistent with our data, but not directly required by the current data. The remaining models assume mild galaxy evolution as formulated by Guiderdoni & Rocca-Volmerange (1987, 1990). The comoving models, with $\epsilon = -1.2$, do not compare favorably with our observations, as they lie above *all* measured correlation amplitudes, including those of early type ('bulge') galaxies, by factors $\gtrsim 2$. These models can be brought into better agreement with our observations by decreasing the correlation length, $r_0$, as might occur if an increasing fraction of galaxies are intrinsically weakly clustered at greater depth, a hypothesis initially suggested by Efstathiou et al. (1991). However, even apart from scaling $r_0$, the relative independence of the comoving models with $I$-magnitude does not follow our combined galaxy sample amplitudes, which show a continuing gradual decline with increasing survey depth.

The correlation amplitudes for bulge-like galaxies are typically a factor of 2–4.5 larger than those for disk-like galaxies. The amplitudes for the disks are not significantly different from those of the combined galaxy samples, for each of the GWS and MDS. We also note that the cross-correlation amplitudes between disks and bulges (open six-pointed stars) are similar to those for the disk and combined galaxy samples. The observed rise in $A_\omega$ toward earlier morphological type is similar in size and magnitude to that found in a similar study we have undertaken using pre-refurbishment WF/PC data (Neuschaefer et al. 1995a), and which is also observed locally by Davis & Geller (1976) and Giovanelli, Haynes & Chincarini (1986).

We finally consider models which allow structures on the scale of galaxy clusters to evolve with time by shrinking relative to comoving coordinates; these models lead to predicted angular correlations which explicitly decline with increasing redshift (see eq.(2)). For such models, $\gamma - \epsilon - 3 \lesssim 0$, i.e. for $\gamma \simeq 1.8$, $\epsilon \gtrsim -1.2$. For $\epsilon = 0$ and $\epsilon > 0$, galaxy clusters are said to maintain their scale and shrink, relative to proper coordinates, respectively. The $\epsilon = 0,1$ models in Figure 3 predict a steady decline of $A_\omega$ with apparent magnitude, consistent with the observed behaviour for the full galaxy correlation amplitudes, although the $\epsilon=1$ model is in best agreement with the full galaxy correlation amplitudes down to $I \lesssim 25$ mag.

### 3.2 Nearest Neighbor Statistics

We have also used nearest neighbor statistics to study the small scale clustering properties of galaxies in the MDS and GWS galaxy fields. Following an approach similar to that used in the analysis of $\omega(\theta)$, the distribution of angular separations between nearest neighbors is compared with that for a field of randomly distributed points occupying the same geometric area. Unlike the $\omega(\theta)$analysis, where all pairs are used, only *nearest* neighbor pairs are counted, thus collapsing the clustering information down to the mean angular separation between neighboring galaxies. In Figure 4, panels (a) and (b), we show the distribution of nearest neighbors in the MDS and GWS fields, respectively. Using the magnitude interval $18 \leq I < 23$ mag, nearest neighbor counts of galaxies classified as disks and bulges are normalized by that of a field of points distributed randomly over the WFPC2 frame. The count distribution for the randomized nearest neighbors is constructed from 20 independent realizations of random points, in which each realization uses the same number of points as in the given galaxy sample. Within the indicated magnitude interval there are typically 40 and 20 disks and bulges per WFPC2 frame. This amounts to $\sim 30$ and 60 nearest neighbor disk and bulge pairs in each of the MDS and GWS samples within $\theta \leq 3''$, corresponding to $\sim 3\%$ and 11% of each of the disk and bulge samples, respectively. As found in the results for the correlation amplitude, disks show weaker apparent clustering than do bulges.

The model curves in Figure 4 are derived from Monte Carlo simulations generated as follows: in a simulated conical volume of space, galaxies are placed in clusters akin to the process of filling a binary tree. The zeroth level of an $N$-level cluster is seeded with two galaxies separated by a distance $d$; at the next level, two child galaxies are placed at opposite ends of a randomly oriented imaginary rod of length $d/\lambda$ centered at each parent galaxy. This process is repeated recursively $N-2$ times, with a reduction of the rod length by the factor $\lambda$ at each level. This construction produces a cluster of $2^N$ objects, with a correlation slope $\gamma = 3 - ln\ 2/ln\ \lambda$; integrated over this conical volume, the correlation amplitude $A_\omega$ is inversely proportional to the number of clusters placed therein. Details of this approach are given in Soneira & Peebles (1978). Effects due to cosmological curvature, galaxy spectral evolution and evolution in the correlation length via exponent $\epsilon$ in eq.(2) are included. To explore the dependence on these parameters, we constructed model universes with values for $\epsilon=-1.2$, 0 and 1, and correlation lengths $r_0=5.5$ and $2.0 h^{-1} Mpc$.

For the MDS sample, the model curves suggest that disks and bulges have similar correlation lengths $\simeq 5 h^{-1} Mpc$, whereas for the GWS sample, disks have substantially fewer close pairs than bulges. Keeping in mind that these results are most sensitive on scales where interactions are important, we speculate that the apparent differences between the GWS and MDS nearest neighbor statistics may relate to differences in maturity of the galaxy merger process. Because of its widely disparate sky coverage, the MDS sample likely represents galaxy pairing statistics for a typical galaxy field, whereas because of known peaks in its redshift distribution ($z = 0.81$, Koo et al. (1996); and $z = 0.98$, Le Fevre et al. (1994)) the GWS field may contain cluster regions which are dynamically more mature, in which potential mergers have already occured by the time of observation.

### 3.3 Evolution of the Galaxy Merger Rate

Estimates of the evolution in the rate of galaxy merging can also be calculated using the nearest neighbor statistics. The nomenclature in recent papers describing evolution in the galaxy two-point correlation function (Efstathiou et al. 1991; Neuschaefer et al. 1995a; Brainerd, Smail & Mould 1995) is useful in this discussion. Except for spatial scales of several kiloparsecs to tens of kiloparsecs, where galaxy merging may be important, we assume that the *spatial* two-point correlation function conforms to a scale-free power law with fixed correlation length $r_0$ and slope $\gamma$:

$$\xi(r, z) = (r/r_0)^{-\gamma} \; (1+z)^{(\epsilon+3)} \tag{6}$$

For the models used in this paper, we adopt $r_0 = 5.5 h^{-1} Mpc$. The exponent $\epsilon$ has values of $-1.2$ and $0$ for galaxy clusters maintaining their shape and scale in comoving and proper coordinates, respectively. For relatively extreme evolution, linear theory predicts density perturbation growth such that $\epsilon \simeq 0.8$ (Yoshii, Peterson & Takahara 1993). In a similar fashion to the presentation of the results in Figure 3, numerous studies of the angular two-point correlation function using deep CCD data have appeared in the literature over the past several years. For $B_J \gtrsim 23$ mag ($I \gtrsim 21$ mag), the correlation amplitude $A_\omega$ is 1.5–3 times smaller than that predicted using models of passive spectral evolution in which galaxy clusters are comoving with the Hubble flow.

Most measurements of $A_\omega$ in deep ground-based surveys have sampled $\omega(\theta)$ on $\sim 10''$ to several arcmin scales, but have not well constrained $\omega(\theta)$ on scales below several arcsec. These surveys have had insufficient angular resolution, limited by atmospheric seeing, to differentiate merging galaxy pairs from starburst regions embedded within single galaxies. Small scale departures of $\omega(\theta)$ from the power law form, caused by an excess or deficit of close pairs on scales of 10–20 $h^{-1} kpc$, provide information on the galaxy pair fraction, $P_f$. The pair fraction is conventionally parameterized as:

$$P_f(z) = P_0(1+z)^m \tag{7}$$

In the foregoing section, we have reported direct observations of the pair fraction, $P_f$, which is related to the galaxy merger rate, $R_{mg}$, and merger timescale, $\tau_{mg}$ as $P_f = R_{mg}\tau_{mg}$. In the next two subsections, we present an upper limit for the evolutionary index, $m$, of the pair fraction, following a procedure which at first excludes effects due to $\omega(\theta)$, and this is then followed by an estimate which includes effects due to $\omega(\theta)$.

### 3.3.1 Upper Limit to the Evolutionary Index, $m$, of the Galaxy Pair Fraction

We can place an upper limit on the evolutionary index, $m$, of the pair fraction using raw galaxy pair counts. Carlberg et al. (1994) used a length scale of 20 $h^{-1} kpc$ to define a "pair" in a statistical sense. Without enquiring in detail into the merging timescale $\tau_{mg}$ (see, for example, Mihos 1995), Carlberg et al. (1994) used a relatively local UGC-based sample to suggest a pairing probability $P_0 = \tau_{mg} R_{mg}^0 \simeq 0.023$ that a given galaxy is a member of a pair with projected linear separation $r < 20$ $h^{-1} kpc$. Down to $V = 22.5$ mag ($I = 21.5$ mag; $z_{med} \simeq 0.4$), they observe the pair fraction to increase to $\tau_{mg} R_{mg} \simeq 10\%$. They interpreted this as evidence for an evolutionary index $m = 2.4 \pm 1.0$, taking $P_0 = 0.03$. Using similar arguments, several other groups have obtained the following indices: (Zepf & Koo, 1989, $m = 4.0 \pm 0.25$; Burkey et al. (1994), $m = 3.5 \pm 0.5$; Yee & Ellingson 1995, $m = 4.0 \pm 1.5$)

There are three types of galaxy pairs: (1) optical doubles, with separations which are large along the line of sight but small on the plane of the sky; (2) galaxy pairs contributing to the hierarchical clustering distribution as described by the spatial two-point correlation function, $\xi(r)$; and (3) galaxy pairs in the process of merging. The latter two types have similar spatial separations but are differentiated by their large and small pairwise velocity differences, respectively. From the sum of these contributions, measurable from the GWS and MDS surveys, we derive an upper limit on the exponent $m$ by using the following expression for the fraction, $p_5$, of galaxies in $\theta < 5''$ pairs which follow a redshift distribution $dN/dz$ corresponding to a given magnitude interval $I_1 - I_2$:

$$p_5(I_1, I_2) = \int_0^z \frac{dN}{dz}\Big|_{I_1, I_2} P_f(z) \, g(z) \, dz \tag{8}$$

where

$$g(z) = \begin{cases} 1, & 5" \geq 20 h^{-1} kpc/D_a(z) \\ 5'' D_a(z)/20 h^{-1} kpc, & 5" \leq 20 h^{-1} kpc/D_a(z) \end{cases} \tag{8.1}$$

$$P(z) = 0.03 \, (1+z)^m \tag{8.2}$$

In eq.(8) $D_a(z)$ is the angular-size distance (Weinberg 1972), and the function $g(z)$ accounts for the declining fraction of physically associated pairs with increasing distance, $D_a(z)$, caused by line-of-sight dilution. The sharp boundary in $g(z)$ lowers the contribution of galaxy pairs with separation $r \leq 20 \ h^{-1} kpc$, since this scale is comparable to the extent of typical galactic disks; within this distance, the identification of neighboring galaxies as distinct sources becomes problematic. Column five of Table 3 gives the fractional excess of galaxy pairs (within $\theta < 5''$) that lies above the predictions for a random distribution, for several magnitude intervals, all of which imply values of $m \leq 2.1$, as given in column six (for $q_0 = 0.05$ these limits rise slightly to $m \leq 2.3$). These values for $m$ result from the assumption of a distribution in $dN/dz$ which has been interpolated from the predictions of Lilly et al. 1995; they are deemed upper limits since they do not account for effects due to $\omega(\theta)$.

### 3.3.2 Best Fit for the Galaxy Pair Fraction Index, $m$

We can derive an estimate for $m$ which accounts for effects due to $\omega(\theta)$ as follows. We define a modified correlation function, $\omega'$, which includes the effect of pair fraction evolution:

$$\omega'(\theta) = \omega(\theta)(1 + \nu_{\mathrm{mg}}(\theta)\zeta) \tag{9}$$

The second term in eq.(9) is used to model the evolution in the pair fraction and includes the following factors: (1) the probability of a merger having a given angular separation, $\nu_{\mathrm{mg}}$, derived from the line of sight integration of the corresponding probability for a merger with a given spatial separation (see eq.(14) below); and (2) a factor $\zeta$ representing the incremental increase in the number of galaxy pairs which arise due to galaxy merging. Such merging causes a decline in the galaxy number density with later cosmological epoch. Thus we express the galaxy number density as $n(z) = n_g^0 \ f(z)$, where $f(z)$ is the number of galaxies which merge into a typical galaxy observed at the present epoch:

$$f(z) = exp\Big( \int_0^z R_{mg}(z) \ \frac{dt(z)}{dz} \ dz \Big) \tag{10}$$

Assuming that the mass-to-light ratio is independent of redshift, the galaxy luminosity function is then:

$$\phi(z, L) = f^2 \phi^* \Big( \frac{fL}{L^*} \Big)^{-\alpha} \tag{11}$$

From Eq. (11) we can formulate two model luminosity functions for the subset of galaxies which undergo mergers: (a) where the merger rate is constant and fixed at the present rate, $R_{mg}^0$:

$$\phi_{mg}^0(z, L) = \tau_{mg}(M) \ R_{mg}^0 \ \phi(z, L) \tag{12.1}$$

and one for which the merger rate varies with $(1 + z)$:

$$\phi_{mg}(z, L) = \tau_{mg}(M) \ R_{mg}(z) \ \phi(z, L) \tag{12.2}$$

with

$$R_{mg}(z) = R_{mg}^0 \ (1 + z)^m \tag{12.3}$$

where $\tau_{mg}$ is the mass-dependent merger time-scale, which can be taken as a constant, to first order, over the range of apparent magnitudes that we are considering. Thus, we can express the enhancement in galaxy numbers due to pair fraction evolution, $1+\zeta$, as the ratio of integrals over the luminosity functions (12.1) and (12.2) (where we use the change of variable, $y = fL/L^*$):

$$1 + \zeta = \frac{\int_0^{z_{max}} \int_{y_{min}(z)}^{y_{max}(z)} (1 + z)^m \ \phi(z, yL^*)\frac{dV}{dz}dz \ dy}{\int_0^{z_{max}} \int_{y_{min}(z)}^{y_{max}(z)} \ \phi(z, yL^*)\frac{dV}{dz}dz \ dy} \tag{13}$$

Because our formulation of $\zeta$ as an enhancement of merging pairs over the value obtained under a constant merger rate the merger time-scale and local merger rate cancel out in our analysis.

The spatial analog to $\nu_{mg}$ can be derived directly as an integral over the galaxy pairwise velocity distribution, $f(v)$:

$$\rho_{mg}(r_p) \propto \int_0^{v_e(r_p)} f(v) \ dv \tag{14}$$

where $v_e = \sqrt{2GM/r_p}$ is the escape velocity parameterized by the pericentric distance $r_p$. To calibrate eq.(14) we use the result from Carlberg et al. (1994), who find that $\nu_{mg}(r_p) \simeq 0.5$ for galaxy pairs brighter than $I \lesssim 21.5$ mag and which have mean separations $r_p \simeq 20h^{-1}kpc$. For eq.(14) we assume an exponential form for the pairwise velocity distribution, as modeled by Davis & Peebles (1983):

$$f(v) = exp(-v/\sigma_v) \qquad (14)$$

where the rms of the pairwise velocity distribution, $\sigma_v \simeq 340$ km/s. The conversion of $\nu_{\rm mg}$ in terms of angle $\theta = r_p/D_a(z)$ is accomplished by the following integration along the line of sight:

$$\nu_{\rm mg}(\theta) = \frac{\int_0^{z_{max}} \frac{dV}{dz}dz \int_{y_{min}(z)}^{y_{max}(z)} dy \ \rho_{\rm mg}(\theta D_a(z))\phi(z, yL^*)}{\int_0^{z_{max}} \frac{dV}{dz}dz \int_{y_{min}(z)}^{y_{max}(z)} dy \ \phi(z, yL^*)} \qquad (15)$$

We determine the index $m$ for evolution of the pair fraction, $P_{\rm f}$, by using eqs(9–15) to minimize $\chi^2$ for our observed correlation functions versus two models for $\omega(\theta)$ with $\epsilon = 1$ and slopes $\delta = -0.8$ and $\delta = -0.6$. The fitting proceeds via modeling $log \ \hat{\omega}(\theta)$ in terms of functional forms linear in the exponent $m$, $log \ \theta$, and the median apparent magnitude of the sample, $I_{\rm med}$. Best fit values for $m$ are given in Table 4 for several magnitude intervals of both surveys, using the fitting interval $1'' \le \theta \le 1'$. All fits are obtained using typically 8–12 $\omega(\theta)$ data points with non-null galaxy-galaxy pairs.

Given the size of the errors in Table 4, there is not strong evidence for a trend of $m$ with apparent magnitude. Because the magnitude intervals partially overlap the individual estimates for $m$ are not completely independent. If this is overlooked we find weighted sums of $m \lesssim 1.2 \pm 0.4$ and $m \simeq 1.8 \pm 0.5$ for slopes $\delta = -0.8$ and $-0.6$, respectively. Excluding the magnitude intervals whose upper limits exceed $I = 24$ mag (corresponding to our nominal 95% completeness limits for all fields, cf. Table 1) we find weighted sums of $m \lesssim 1.5 \pm 0.6$ and $m \simeq 1.9 \pm 0.7$ for the same correlation slopes. Both of these results indicate substantially less evolution in the pair fraction than that claimed by Zepf & Koo (1989), Burkey et al. (1994), or Carlberg et al. (1994), who find excess galaxy pairs indicating $m \gtrsim 2.5$. But we find evidence for slightly more evolution than Woods et al. (1995), who find no significant excess of galaxy pairs above that for a randomly distributed sample. This difference is somewhat mitigated by the possible flattening of $\omega(\theta)$ toward larger redshift suggested by NW95.

The modest evolution of the pair fraction, in combination with the correlation amplitudes for disks and bulges discussed in §3, leads us to consider the following possibility. The classified disks of the MDS and GWS samples have low correlation amplitudes and may overlap substantially with the excess population of dwarf galaxies observed in deep $B$-band surveys (e.g., Tyson 1988), which are known to have very low values of $A_\omega$ (Efstathiou et al. 1991; NW95). This identification suggests that the fate of a large fraction of the weakly clustered disk galaxies (prominent in the $B$-band) is that they fade or dissipate by the present epoch, but that they do not undergo wholescale merging.

## 4. GALAXY PAIRS WITH SEPARATIONS $\theta \le 3''$

We have undertaken a systematic visual inspection of close galaxy pairs from the GWS and MDS surveys. We selected galaxy pairs from objects classified as disks, bulges or generic galaxies brighter than $I \le 23$ mag and having separations $\theta \le 3''$, thus accruing 267 galaxy pairs/groups in the combined MDS and GWS surveys. In this effort two of us (LWN and MI) visually inspected the combined pair sample independently, and categorized the pairs among the following classes: (a) 'apparently physically associated', used to describe galaxy pairs which have surface brightness features linking the neighboring galaxies, but which are morphologically undisturbed; (b) 'strongly interacting/merging', used to describe pairs with strongly disturbed morphology to an extent that outer isophotal contours are difficult to identify with one or the other galaxy; (c) 'star-forming regions', used to describe groupings of several compact, high surface brightness features enveloped in an elongated low surface brightness structure suggestive of a single underlying galaxy having either disk or irregular morphology; or (d) physically unassociated. In this effort we each compiled lists of $\sim 100$ pairs which were judged to be physically linked as described by categories (a–c); 48 of the physically associated pair candidates are common to both observers' lists. In a second stage we examined all cases of disagreement where only one of us identified a given pair as *not* physically associated, upon which we reapplied the same categorization as used for the initial galaxy pair sample. As a result of this inspection we identified 86 pairs ($\simeq 32\%$) which showed evidence for physical association. This is somewhat lower than the value of $\omega(\theta) \sim 45\%$ within $3''$, I$\leq 23$ mag, obtained as a power law extrapolation from $1'$ scale, using log $A_\omega = -1.5$, $\delta = -0.8$ (see Table 2). Plates 1–2 show montages of candidate galaxy pairs or groups which are physically associated and come from the combined MDS and GWS surveys. We note in passing that one or both of us (LWN,MI) finds 32 of the 86 physically-associated pairs to appear as starburst regions within single galaxies. Thus, for the excess numbers of close pairs to contribute to a power law form for $\omega(\theta)$ down to arcsec scales (excluding excess pairs associated with pair fraction evolution), more than half of the pairs which are *actually* physically associated reveal no morphological evidence for such association.

Although our somewhat subjective selection criteria makes it difficult to provide a quantifiable analysis of the morphological properties for the galaxy pairs/groups shown in Plates 1–2, we make several observations on the nature of the galaxy interactions seen in our data. Roughly a third of our galaxy pairs have low surface brightness features which generally extend toward their presumed interacting neighbor. About one in five pairs shows evidence for extensions possibly caused by tidal stripping; one in three pairs has compact knots within extended low surface brightness envelopes; one in ten pairs has a very low surface brightness galaxy as a pair member; and one in twenty shows actual internal disturbances suggestive of active merger processes. Finally, one in ten pairs is ambiguous regarding whether their constituents had a prior history as separate entities, or are star-forming regions within a single galaxy.

Figure 5 shows the distribution of magnitude differences within our galaxy pair candidates which are classed as physically associated. In the same figure we have plotted scaled distributions for the full sample of $\theta \leq 3''$ galaxy pairs (short dashed histogram), for a spatially random galaxy sample with galaxy counts consistent with Figure 1 (dot-dashed histogram), and for the $\epsilon = 1$ cluster simulation discussed in §3.2. The distribution for the interacting pair sample is not substantially different from that for the complete sample of close galaxy pairs. In turn, both these distributions are consistent with that for the random galaxy simulation. We have compared the $(V-I)$ color distribution of the physically associated pair candidates with that of the full pair sample and find no systematic trend. However, this lack of correlation between morphological disturbance and $(V-I)$ color is most probably due to the limited color baseline between the wide $F606W$ and $F814W$ passbands.

We emphasize that the upper limit $I_{lim} \leq 23$ mag used in Figures 5 and 6 corresponds to practically 100% completeness, even for generic galaxies of the lowest surface brightness, as is apparent from Figure 1. Moreover, all the galaxy subsamples used to determine $m$ are at least two magnitudes wide (Table 4) and so are unlikely to miss a substantial fraction of galaxy pairs due to greatly differing magnitude differences. Thus, the values for $m$ determined in Table 4 are largely unaffected by selection effects for $I \leq 23$ mag and are valid for fainter magnitudes *if* the distribution of magnitude differences for close galaxy pairs with $I \geq 23$ is similar to that shown in Figure 5. However, if the magnitude difference distribution for close galaxy pairs is substantially broader for $I \geq 23$ mag than found in Figure 5 this would amount to substantial population close galaxy pairs which would go undetected, possibly causing an underestimate for $m$.

## 5. DISCUSSION

Depending on whether or not the correlation slope flattens with increasing look-back time, the results of §3 suggest that galaxy merging at moderate redshift has occurred at a moderately or only marginally higher rate than at present. The results of §4 are more suggestive of marginal evolution in the galaxy pair fraction. In this section we enquire as to how the pair fraction we observe can be reconciled with the latest number counts and redshift distributions (e.g., Broadhurst et al. 1992). The argument for a substantial amount of galaxy merging has been based on the combination of galaxy number counts in $B_J$ and $K$, together with the galaxy redshift distributions. The $B_J$ number counts suggest an excess of galaxies relative to a 'no evolution' model, an excess which is not apparent in either the $K$-band counts or redshift distributions. There have been several groups who have measured the galaxy redshift distributions down to $B_J \lesssim 24$ mag ($I \lesssim 22$ mag). The spectroscopic surveys of Broadhurst, Ellis & Shanks (1989) and Colless et al. (1990, 1993) limited to $B_J \leq 22.5$ mag ($I \lesssim 20.5$ mag), yeild a sample with a median redshift $z_{med} \simeq 0.3$, whereas the spectroscopic survey of Glazebrook et al. (1995) limited to $B_J \leq 24$ mag ($I \lesssim 22$ mag) yeilds a sample with a median redshift of $z_{med} \simeq 0.5$. Depending on the flux limit, the no evolution model underestimates these observers' redshift distributions by factors of 1–3, but the no evolution model does fairly trace the *shape* of the observed redshift distributions. The merging model proposed by Broadhurst et al. (1992), such that a typical present day galaxy is formed from $\sim 4$–6 subunits existing at $z = 1$, is able to simultaneously account for the above redshift distributions, and number counts in $B_J$ and $K$ passbands.

From pre-refurbishment MDS WF/PC data, Im et al. (1995a) found the predicted galaxy sizes of the Broadhurst et al. (1995) merging model to be marginally consistent with the observed (total) size distribution down to $I \lesssim 22$ mag. In a more detailed study on this issue, however, Im et al. (1995b) considered the merging model as a means to interpret the E/S0 galaxy size distribution from the MDS WFPC2 data. They found an upper limit to the pair fraction exponent $m \lesssim 1$ in the flux interval $20 \leq I \leq 22$ mag; larger values for $m$ led to a model median scale-length significantly smaller than that observed. We note there is the potential for overestimating galaxy sizes, as the data in Im et al. 's study approaches the resolution-limited regime. However, Monte Carlo tests on a preliminary version of the model-fitting software used by Im et al. (1995b) (see Ratnatunga et al. 1996) showed no biases for half-light radii $\gtrsim 0\rlap{.}''1$, much smaller than the median $r_{hl} \gtrsim 0\rlap{.}''4$ Im et al. found for E/S0 galaxies in the range $20 \leq I \leq 22$ mag. Moreover, further study of data in the GWS galaxy sample confirms Im et al. 's E/S0 median size estimate, using a revised version of the software by Ratnatunga et al. (1996) which can estimate galaxy sizes in an unbiased manner for $r_{hl} \gtrsim 0''\!.03$. Because of the lack of evidence for pair fraction evolution in the present study our results suggest that galaxies later than E/S0 have a similar, passive, history of merging.

The large spectroscopic survey of $\sim 700$ galaxies limited to $I_{ab} \leq 22.5$ mag ($I \lesssim 22$ mag) of Lilly et al. (1995), provides a more detailed picture of galaxy evolution which can explain many of the effects formerly attributed to the proposed galaxy merger process. The redshift distribution for this survey has a median $z_{med} \simeq 0.55$, as well as a population of blue galaxies at $z \leq 0.2$ significantly in excess of the local luminosity function of Loveday et al. (1992). Exploiting the large sample size and the relatively optimal sampling in the $I$-band (which avoids the substantial bias against detecting intrinsically faint red galaxies which occurs in the $B_J$ passband), Lilly et al. (1995) use their survey to measure significant color-dependent evolution of the galaxy luminosity function. Namely, they find that beyond $z \gtrsim 0.5$, galaxies bluer than present day Sbc brighten by roughly 1 magnitude, whereas redder galaxies brighten by no more than a few tenths of a magnitude. This observation, combined with the low redshift population discovered within their own sample, suggests that the excess in the blue number counts is largely a consequence of color-dependent luminosity evolution. If so, the large numbers of late-type and irregular galaxies recently observed by Glazebrook et al. 1995 and Driver et al. 1995 may be a manifestation of both Lilly *et al.'s* proposed color-dependent luminosity evolution as well as the weak correlations we observe in Figure 3.

We should point out that the results in §3 are consistent with previous results obtained for the analysis WF/PC data applied to the study of clustering versus morphological type (Neuschaefer et al. 1995). Morphology segregation has been observed by several groups at brighter magnitudes including Davis & Geller (1976), Giovanelli, Haynes & Chincarini (1986) and Loveday et al. (1995), all of whom find that the amplitude of early-type galaxies is a factor of 3–5 larger than for late-type galaxies at a scale of $1h^{-1}Mpc$. Moreover, within $1h^{-1}Mpc$, these same researchers all find that the slope of $\omega(\theta)$ (or equivalently $\xi(r)$) is steeper for early- than late-type galaxies.

## 6. CONCLUSIONS

We have examined the small- and intermediate-scale angular correlations of faint galaxies observed using WFPC2 on HST. These high resolution images allow the study of clustering properties versus galaxy morphology out to median redshifts $z_{med} \lesssim 1$. In the present galaxy correlation study we find the following:

(i) The correlation amplitude, $A_\omega$, for combined disk plus bulge galaxy samples shows apparently strong evolution in the sense of increasing moderately rapidly with *decreasing* median redshift. The viable explanations are: (a) mild galaxy luminosity evolution in a low-density universe ($q_0 = 0.05$) with strong clustering evolution ($\epsilon \gtrsim 0$), wherein galaxy-cluster-scale structures are gradually collapsing in proper coordinates; or (b) the onset of a weakly clustered population of dwarf galaxies which make their appearance for $I \gtrsim 22$ mag (Im et al. 1995a); or (c) possible evolution in the correlation slope which can explain the apparently strong decline in $A_\omega$ but which is only marginally consistent with our observations of $\omega(\theta)$ vs $\theta$.

(ii) The correlation amplitude for bulge-dominated galaxies is 2–4 times larger than that for disk-dominated galaxies, similar to what is observed locally in rich cluster environments.

This result is observed for both the contiguous Groth-Westphal Survey field and the disjoint, randomly distributed MDS survey. Since the MDS sample is comprised of randomly pointed fields, its clustering properties are fairly representative of the field. These results indicate that morphological segregation is fairly independent of galaxy space density.

(iii) In contrast to the substantial evolution in $A_\omega$, the correlation function $\omega(\theta)$ for combined disk plus bulge galaxy samples undergoes negligible steepening at small angular scales. We parameterize the evolution of the galaxy pair fraction $P_{\rm f} \propto (1+z)^m$, and find an upper limit of $m \lesssim 1$, using the observed fraction of galaxies within $5''$ and not accounting for effects due to $\omega(\theta)$. By modeling $\omega(\theta)$ to include effects due to $P_{\rm f}$, we find modest evolution with $m = 1.2 \pm 0.4$. Taking into consideration the results from (i) and (ii) above, then if a weakly clustered population of dwarf galaxies is mostly responsible for the excess galaxy counts in $B_{\rm J}$, the relatively low value for $m$ suggests that the majority of these galaxies have not merged with neighbors but have either faded or dissipated by the present epoch.

(iv) The fraction of galaxy pairs with separations $\theta \lesssim 3''$ (or $\lesssim 9h^{-1}kpc$ at the median redshift $z_{\rm med} \lesssim 0.5$ of our galaxy sample), and with visible evidence for physical association, is two thirds that needed to account for an extrapolation of $\omega(\theta)$ down to arcsecond scales, assuming a constant correlation slope $\delta = -0.8$. The $(V-I)$ color and $I$-magnitude difference distributions for the galaxies in these pairs is similar to that for a randomly spatially distributed galaxy sample.

## ACKNOWLEDGEMENTS

This work was supported by NASA/HST grants GO-2684-0X-87A and GO-3917-0X-91A from STScI, which is operated by AURA, Inc., under NASA contract NAS5-26555. We acknowledge several useful comments from Rogier Windhorst.

## TABLE CAPTIONS

**Table 1.** Col. (1): Survey specifier; Col (2)–(3): Galactic $(l, b)$; Col (4): Number of exposures in $I$-band stacked image; Col (5): Number of fields within 0.5 diameter region; Col (6): Total integration time; Col (7): Sky flux in units of ADU per hour; Col (8): 95% completeness limit for galaxies with $r_{\rm hl} \lesssim 0\overset{''}{.}8$.

**Table 2.** Col (1): Median $I$-magnitude; Col (2): $I$-magnitude interval; Col (3): Survey specifier; Col (4)–(5): best-fit and $1\sigma$ error estimate of correlation amplitude, $log\ A_\omega$; Col (6)–(7): Numbers of galaxies and galaxy pairs in the survey region, respectively; Col (8): Galaxy type selected for sample.

**Table 3.** Col. (1): I-magnitude interval; Col. (2): Number of galaxies in the sample; Col. (3): Number of galaxy-pairs in the sample; Col (4): Number of galaxy pairs in excess of that expected from a randomly distributed set of points; Col (5): Fractional excess in the galaxy pairs above random statistics; Col (6): Upper limit to the evolutionary exponent, $m$, in the galaxy pair fraction.

**Table 4.** Col. (1): I-magnitude interval; Col. (2): Survey specifier; Cols. (3)–(6): best fit, error and reduced $\chi^2$ for the evolutionary index, $m$, in the galaxy pair fraction, assuming a correlation slope $\delta = -0.8$; Cols. (7)–(10): best fit, error and reduced $\chi^2$ for $m$, assuming a correlation slope $\delta = -0.6$.

## FIGURE CAPTIONS

**Figure 1.** $I$-band number counts for galaxies classified as generic (dot-dashed), disk (solid) and bulge (dashed) from the GWS survey.

**Figure 2.** Angular correlation functions of galaxy bulges from the Groth-Westphal Survey. Baselines with slope $\delta = -0.8$ are drawn to guide the eye. Poisson error estimates are plotted, with downward pointing arrows indicating a lower limit for $\omega(\theta) \lesssim 0$.

**Figure 3.** Correlation amplitudes, $A_\omega$, for disks (filled squares), bulges (filled hexagons) and combined disk, bulge and generic galaxy samples (open circles) for the GWS and MDS samples in upper and lower panels, respectively. Also plotted are the cross-correlation amplitudes for both samples (open six-sided stars), and previous estimates of the generic galaxy correlation amplitudes from the ground (Neuschaefer & Windhorst 1995; open triangles) and pre-refurbishment WF/PC data (Neuschaefer et al. 1995; five-pointed skeletal points). Model curves represent clustering evolution with $r_0 = 5.5 h^{-1} Mpc$, where galaxy clusters maintain their scale in the comoving frame ($\epsilon = -1.2$: short-dashed and dot-long-dashed lines); where galaxy clusters maintain their scale in the proper coordinate frame but shrink relative to the comoving frame ($\epsilon = 0$: long dashed line); where clusters

shrink in scale in the proper coordinate frame ($\epsilon$=1: solid line); and finally a model with slope evolution, for which ($\epsilon$=0, $\gamma$=1.8 $(1 + z)^{-0.2}$; dotted line).

**Figure 4.**   (a) MDS nearest neighbor counts versus angular separation, $\theta$, normalized by the nearest neighbor counts for a random sample. Open triangles: disk/disk pairs; open circles: bulge/bulge pairs; open squares: disk/bulge pairs.   (b) The same statistics are repeated for the GWS survey. Clustering models are shown with $q_0$=0.05: the comoving model with $\epsilon$=−1.3 (dashed curve); and the strong clustering evolution model with $\epsilon$=1.0, (dotted curve).

**Figure 5.**   The distribution of magnitude differences within galaxy pairs with pair members brighter than $I \leq 23$ mag and with pair separations $\theta \leq 3''$.   The apparently physically associated pair sample is shown as the solid histogram; the total pair sample is shown as the dashed histogram; and the random galaxy sample is shown as the long dashed curve. The results of the simulation with a clustering evolution exponent $\epsilon$=1 is shown as the dot-dashed curve.

**Plates 1−2.**   Montages of apparently interacting galaxies from the MDS and GWS surveys.

**TABLE 2**

$I-$band power law fits for $\omega(\theta)$

| $I_{med}$ [mag] | $I$-Range [mag] | Survey | $log\ A_{\omega}$ mean | error | $N_{gal}$ | $N_{pair}$ | Galaxy Type |
|---|---|---|---|---|---|---|---|
| 19.3 | $18.0-20.0$ | GWS | $-0.76$ | 0.10 | 115 | 6555 | galaxy |
| 20.3 | $18.0-21.0$ | GWS | $-1.18$ | 0.07 | 342 | 58311 | galaxy |
| 21.3 | $18.0-22.0$ | GWS | $-1.20$ | 0.09 | 798 | 318003 | galaxy |
| 22.3 | $18.0-23.0$ | GWS | $-1.42$ | 0.11 | 1622 | 1314401 | galaxy |
| 23.7 | $23.0-24.0$ | GWS | $-1.68$ | 0.16 | 1573 | 1236378 | galaxy |
| 24.7 | $24.0-25.0$ | GWS | $-1.83$ | 0.10 | 2978 | 4432753 | galaxy |
| 20.9 | $18.0-22.0$ | GWS | $-1.63$ | 0.32 | 439 | 96141 | disk |
| 21.8 | $18.0-23.0$ | GWS | $-1.65$ | 0.24 | 1011 | 510555 | disk |
| 22.3 | $21.0-23.0$ | GWS | $-1.67$ | 0.27 | 859 | 368511 | disk |
| 23.3 | $22.0-24.0$ | GWS | $-1.57$ | 0.17 | 1470 | 1079715 | disk |
| 24.3 | $24.0-24.5$ | GWS | $-1.61$ | 0.34 | 316 | 49770 | disk |
| 20.9 | $18.0-22.0$ | GWS | $-1.03$ | 0.09 | 274 | 37401 | bulge |
| 21.8 | $18.0-23.0$ | GWS | $-1.04$ | 0.06 | 435 | 94395 | bulge |
| 22.3 | $21.0-23.0$ | GWS | $-1.08$ | 0.11 | 281 | 39340 | bulge |
| 23.3 | $22.0-24.0$ | GWS | $-1.18$ | 0.15 | 295 | 43365 | bulge |
| 21.1 | $18.0-22.0$ | GWS | $-1.50$ | 0.25 | $-$ | 120286 | disk/bulge |
| 22.9 | $22.0-23.5$ | GWS | $-1.81$ | 0.35 | $-$ | 242760 | disk/bulge |
| 19.3 | $18.0-20.0$ | MDS | $-0.72$ | 0.13 | 110 | 217 | galaxy |
| 20.3 | $18.0-21.0$ | MDS | $-1.09$ | 0.22 | 283 | 1426 | galaxy |
| 21.3 | $20.0-22.0$ | MDS | $-1.40$ | 0.09 | 645 | 7483 | galaxy |
| 22.3 | $18.0-23.0$ | MDS | $-1.62$ | 0.08 | 1591 | 45199 | galaxy |
| 23.7 | $23.0-24.0$ | MDS | $-1.92$ | 0.12 | 1791 | 57306 | galaxy |
| 24.7 | $24.0-25.0$ | MDS | $-2.08$ | 0.11 | 3111 | 172856 | galaxy |
| 21.3 | $20.0-22.0$ | MDS | $-1.24$ | 0.32 | 375 | 2659 | disk |
| 22.3 | $21.0-23.0$ | MDS | $-1.61$ | 0.41 | 818 | 12444 | disk |
| 23.3 | $23.0-23.5$ | MDS | $-1.76$ | 0.53 | 468 | 4094 | disk |
| 24.1 | $23.5-24.5$ | MDS | $-1.67$ | 0.40 | 965 | 25934 | disk |
| 21.3 | $20.0-22.0$ | MDS | $-1.02$ | 0.26 | 228 | 1079 | bulge |
| 22.3 | $21.0-23.0$ | MDS | $-1.22$ | 0.32 | 323 | 1987 | bulge |
| 23.3 | $23.0-23.5$ | MDS | $-0.96$ | 0.31 | 87 | 199 | bulge |
| 21.1 | $18.0-22.0$ | MDS | $-1.43$ | 0.39 | $-$ | 3441 | disk/bulge |
| 22.9 | $22.0-23.5$ | MDS | $-1.60$ | 0.41 | $-$ | 8994 | disk/bulge |

**TABLE 3**

Excess Pair Counts within $1'' \leq \theta \leq 5''$, GWS survey

| $I$-mag range | $N_{\mathrm{gal}}$ | $N_{\mathrm{pair}}$ | $N_{\mathrm{ex}}$ | $f_{\mathrm{ex}}$ | $m_{\mathrm{ulim}}$ |
|---|---|---|---|---|---|
| 19.0−23.5 | 1776 | 850 | 125 | 0.15 | 2.6 |
| 21.0−24.5 | 3140 | 2441 | 217 | 0.09 | 2.7 |
| 22.5−25.5 | 5557 | 7556 | 648 | 0.09 | 3.1 |
| 24.0−25.5 | 3400 | 3368 | 320 | 0.10 | 2.6 |

**TABLE 4**

Best-Fit Galaxy Pair Fraction Index, $m$

| $I$-Range [mag] | Survey | $\delta = -0.8$ | | | $\delta = -0.6$ | | |
|---|---|---|---|---|---|---|---|
| | | $m$ | $\delta m$ | $\chi^2$ | $m$ | $\delta m$ | $\chi^2$ |
| 18−21 | GWS | 2.3 | 2.0 | 1.2 | 2.8 | 2.2 | 1.3 |
| 20−22 | GWS | 1.1 | 2.3 | 1.1 | 1.4 | 2.4 | 1.3 |
| 21−23 | GWS | 1.9 | 1.3 | 1.1 | 2.3 | 1.4 | 1.2 |
| 22−24 | GWS | 1.0 | 1.2 | 1.5 | 1.5 | 1.4 | 1.6 |
| 23−25 | GWS | 2.4 | 1.4 | 1.3 | 2.9 | 1.4 | 1.5 |
| 24−25 | GWS | 1.8 | 1.2 | 1.1 | 2.2 | 1.2 | 1.3 |
| 19−21 | MDS | 1.0 | 3.1 | 1.8 | 1.5 | 3.2 | 1.9 |
| 18−22 | MDS | 1.9 | 1.4 | 1.3 | 2.5 | 1.9 | 1.5 |
| 21−23 | MDS | 2.6 | 2.0 | 1.2 | 2.9 | 2.1 | 1.3 |
| 22−24 | MDS | −0.1 | 1.7 | 1.1 | 0.3 | 2.1 | 1.3 |
| 23−25 | MDS | 0.1 | 1.3 | 1.1 | 0.6 | 1.4 | 1.2 |
| 24−25 | MDS | 0.7 | 1.3 | 0.9 | 1.3 | 1.4 | 1.2 |